\documentclass[useAMS,usenatbib]{mn2e}
\usepackage{graphics,graphicx,epsfig,psfig}
\usepackage[normalem]{ulem}
\usepackage[dvipsnames]{color}
\usepackage{soul,xcolor}
\usepackage{amsmath, amssymb}
\usepackage[]{inputenc,amssymb}
\usepackage{booktabs,caption}
\usepackage{amsmath}

\setstcolor{red}

\def \be{\begin{equation}}
\def \ee{\end{equation}}
\def \ba{\begin{eqnarray}}
\def \ea{\end{eqnarray}}
\def \etal{{et al.}}

\newcommand{\sfr}{M$_{\odot}$ yr$^{-1}$}

\newcommand{\ucre}{u$_{\rm{CRe}}$}
\newcommand{\ugas}{u$_{\rm{gas}}$}
\newcommand{\ergps}{erg s$^{-1}$}

\definecolor{webgreen}{rgb}{0,.5,0}
\definecolor{webbrown}{rgb}{.6,0,0}
\usepackage[%
    colorlinks = true,%
    linkcolor = blue,%
    urlcolor  = blue,%
    citecolor = webgreen,%
    anchorcolor = blue]{hyperref}

\newcommand{\ufhref}[3][blue]{\href{#2}{\color{#1}{#3}}}%

\def \ergps{erg s$^{-3}$}

\def \etal{\textit{et. al. }}

\definecolor{webgreen}{rgb}{0,.5,0}
\definecolor{webbrown}{rgb}{.6,0,0}

\usepackage[T1]{fontenc}
\usepackage{ae,aecompl}

\usepackage{newtxtext,newtxmath}
\title[Radio halos of star forming galaxies]{Radio halos of star forming galaxies}

\author[Vijayan \etal]
{
\parbox{\textwidth}{Aditi Vijayan$^{1,2}$\thanks{Email: aditiv@rri.res.in},
 Biman B. Nath$^{1}$, 
 Prateek Sharma$^2$ and
 Yuri Shchekinov$^3$} \vspace{0.4cm}\\
\parbox{\textwidth}{ 
$^{1}$Raman Research Institute, Bangalore, 500080, India\\
$^{2}$Department of Physics and Joint Astronomy Programme and Department of Physics, Indian Institute of Science, Bangalore 560012, India\\
$^3$ Lebedev Physical Institute of Russian Academy of Sciences, ASC, Moscow 117997, Russia\\
}
}
\date{Last updated 2015 May 22; in original form 2013 September 5}

\pubyear{2015}
\begin{document}
\maketitle
\label{firstpage}
\begin{abstract}
We study the synchrotron radio emission from extra-planar regions of star forming galaxies. We use ideal magneto-hydrodynamical (MHD) simulations of a rotating Milky Way-type disk galaxy with distributed star formation sites for three star formation rates (SFRs) ($0.3$, $3$, $30$ \sfr). From our simulations, we see emergence of galactic-scale magnetised outflows, carrying gas from the disk. We compare the morphology of the outflowing gas with hydrodynamic (HD) simulations. We look at the spatial distribution of magnetic field in the outflows. Assuming that a certain fraction of gas energy density is converted into cosmic ray energy density, and using information about the magnetic field, we obtain synchrotron emissivity throughout the simulation domain.
We generate the surface brightness maps at a frequency of $1.4$ GHz. The outflows are more extended in the vertical direction than radial and hence have an oblate shape. We further find that the matter right behind the outer shock, shines brighter in these maps than that above or below. To understand whether this feature can be observed, we produce vertical intensity profiles.
We convolve the vertical intensity profile with the typical beam sizes of radio telescopes, for a galaxy located at
$10$ Mpc (similar to NGC $891$) in order to estimate the radio scale height to compare with observations. We find that for our SFRs this feature will lie below the RMS noise limit of instruments.
The radio scale height is found to be $\sim 300-1200$ pc
, depending on the resolution of the telescope. We relate the advection speed of the outer shock with the surface density of star formation as ${\rm v}_{\rm adv} \propto \Sigma_{\rm SFR}^{0.3}$, which is consistent with earlier observations and analytical estimates.

\end{abstract}

\section{Introduction}
Morphology of galaxies in different wavebands provide us with clues to various physical processes occurring in 
a galaxy, and therefore helps us to understand different aspects of galactic evolution. The physical processes such as outflows, exchange of mass and momentum between the star forming disk and the circum-galactic medium, unleashed 
by star formation activity in a galaxy have far reaching consequences, ranging
from regulation of star formation rate (SFR) in a galaxy \citep{Krumholz2012, Lim2018, Sarkar2019}, to influencing the physical state of the circum-galactic medium \citep{Tumlinson2017} and  enrichment of the intergalactic medium (e.g., \citealp{Oppenheimer2006}). Since these physical processes are complex in nature, and are mediated by multiple factors (e.g., interaction between various gas phases, cooling), it is important to compare theoretical predictions with observations in different wavebands in order to ensure the validity of models of stellar feedback. In this regard, results of hydrodynamical simulations have been compared with optical \citep{Cooper2008}, X-ray \citep{Suchkov1996, Cooper2009, Vijayan2018}, and molecular lines \citep{Roy2016}.

Radio emission from star formation in galaxies has long been an important field of study \citep{Condon1992}. Synchrotron emission produced by cosmic ray electrons, accelerated by supernovae, dominates the radio luminosity ($\nu < 10$ GHz) of star forming galaxies (remaining small fraction arising from free-free emission in HII regions). Observations of edge-on spiral galaxies have shown that the synchrotron emission can reach vertical heights of several kpc \citep{Hummel1989, Hummel1991, Dahlem1995, Dahlem2006, Irwin1999,Irwin2012}. The vertical extent of the radio halo, referred to as the radio scale height, depends on the magnetic field configuration and cosmic ray transport properties. It is smaller than the magnetic scale height \citep{Beck2009}.
The  radio scale height is also related to advection of cosmic rays, and therefore, possibly related to the star formation process, which may dictate the convective speed. Cosmic rays, thus, can potentially trace galactic outflows, and the study of radio halos is important in the overall understanding of physics of outflows.

The morphology of radio halos of star forming galaxies has become important from another quarter as well. Recent balloon measurements with ARCADE-2 have shown the existence of a bright Galactic high latitude diffuse radio emission, referred to as the monopole \citep{Fixsen2011, Kogut2011}, confirming early tentative measurements (e.g.,  \citealp{Costain1960}). The magnitude of the monopole crucially depends on the assumed radio halo morphology of our Galaxy. Explaining this excess brightness (with brightness temperature of order $\sim 1$ K at $1.4$ GHz) with an extragalactic component is difficult because the excess does not have a far-infrared counterpart as predicted by the observed strong radio/far-IR correlation in galaxies. \citealp{Biermann2014} have invoked models of emission from shocks due to primordial massive stars at cosmological distances to understand this excess. To avoid such models, \citealp{Subrahmanyan2013} tried to explain the emission by invoking a spherical radio halo of our Galaxy extending up to $\sim 15$ kpc. But this has been doubted by \citealp{Singal2015} by calling attention to the fact that observed radio morphology of edge-on galaxies do not show such a spherical geometry. 

Observations of radio halos of edge-on galaxies give a number of clues to their connection with star formation activity in the disk. For example, \citealp{Dahlem1995} used observation of five edge-on galaxies to establish that a critical star formation rate surface density, ($\dot{E_A} > 10^{-4}$ erg s$^{-1}$ cm$^{-2}$ ) is required to support an extended halo. Assuming a Kroupa initial mass function, $10^{51}$ erg per supernovae and that a fraction $\sim 0.3$ thermalisation of energy deposited, translates into a star formation rate surface density of $\sim  10^{-2}$ M$_\odot$ yr$^{-1}$ kpc$^{-2}$. Incidentally, this threshold is an order of magnitude lower than that of launching a galactic outflow \citep{Lehnert1996}. \citealp{Heesen2016} found that radio halos exist for galaxies with a star formation rate density of $\ge 10^{-3}$ M$_{\odot}$ yr$^{-1}$ kpc$^{-2}$. \citealp{Roy2013} and \citealp{Vasiliev2019} explain these thresholds with the help of hydrodynamical simulations. An explicit connection to the radio halo would require information about the galactic magnetic field.

\citealp{Dahlem1995} found the morphology of radio halos to be oval-shaped, with the radial extent of the halo at the radio scale height being somewhat smaller than that along the mid plane of the disk. The radial extent of the halo in the disk plane was further found to be larger (by a few kpc, on an average) than that of star formation. They explained this difference by invoking cosmic ray diffusion, with a value of diffusion coefficient $D \sim10^{28}\hbox{--} 10^{29}$ cm$^2$ s$^{-1}$, consistent with cosmic ray phenomenology. Interestingly, \citealp{Dahlem2006} found the radial extent at a height to be comparable to the star formation extent along the disk. Clearly, the extra-planar radio emission is related to the distribution of star formation sites in the disk. This has been confirmed by \citealp{Krause2018}, who observed a sample of $13$ edge-on spiral galaxies and found that the radio scale height depends on the extent of radio emission along the plane, but not on SFR. The star formation rate surface density in their sample of galaxies was low ($\le 7 \times 10^{-3}$ M$_\odot$ yr$^{-1}$ kpc$^{-2}$).

Due to uncertainties in obtaining the SFR in \citealp{Dahlem1995}, the correlation between extent of radio halos and SFR was not replicated by \citealp{Irwin1999}. \citealp{Irwin1999} attribute this to difference in time scales over which star formation ($\sim 10^7$ yr) takes place and that over which radio halos can persist ($\sim 10^8$ yr, because of cosmic ray electron (CRe) lifetime, which is given by $\sim 0.8 \times 10^8$ yr $(\nu/1.4 \, {\rm GHz})^{-0.5} \, (B/ \mu{\rm G})^{-1.5}$). Further, \citealp{Irwin1999} observed discrete features above the star forming regions in the disk characterised by spectral flattening, which they ascribe to winds launched by star formation process. 
In their spectral maps of galaxies, they noticed a steepening of the spectral index (i. e. it becomes more negative) away from the plane of the disk along the minor axis. They explained this by arguing that CRes that populate the halo would have lost energy in the disk before being transported to the halo \citep{Duric1998}. 

There are two time scales involved here- 1. the time scale of cosmic ray transport to extra-planar heights, which can occur either through diffusion or advection in case of an outflow, and 2. the radiation loss time scale of electrons. As mentioned above, the radiation loss time scale of electrons radiating at $1.4$ GHz is of order $\sim 80$ Myr for $\mu$G magnetic field. In comparison, the diffusion time scale is {$l^2/2D\sim 180$ Myr} for a height of $3$ kpc and diffusion coefficient of $10^{28}$ cm$^2$ s$^{-1}$. Advection time scale to reach the same height would be of order $\sim 30$ Myr for a wind speed of $100$ km s$^{-1}$. Therefore, a spectral steepening with height would correspond to the case of large magnetic field (short cooling time scale) and low wind speed. Confirming this, \cite{Duric1998} mentioned that morphology of radio halos with spectral flattening at extra-planar regions did indicate the presence of winds, and had features of `chimneys' through which disk material had broken out. This was confirmed by high resolution observations of \citealp{Irwin2000}, with the presence of discrete features in the radio halos.
 
These ideas were confirmed by \citealp{Heesen2016}, who used 1-D transport model to understand how advection and diffusion modify the spectral index. They also found that while advection causes the spectral index to steepen gradually, in a diffusion dominated scenario the spectral index hardly steepens within one scale height and steepens rapidly at larger heights. Recently \citealp{Heesen2018} used 1-D cosmic ray transport equations, of advection and diffusion, in order to model the vertical intensity and spectral profile, fitting to radio continuum maps of $12$ nearby galaxies. They found that an advection speed scaling, ${\rm v}_{\rm adv} \sim \Sigma_{\rm SFR}^{0.3}$, can explain the profiles. In the study by \citealp{Krause2018} it was inferred that convection plays an important role in transporting cosmic rays to extra-planar heights. 

\begin{figure*}
	\includegraphics[width=\textwidth]{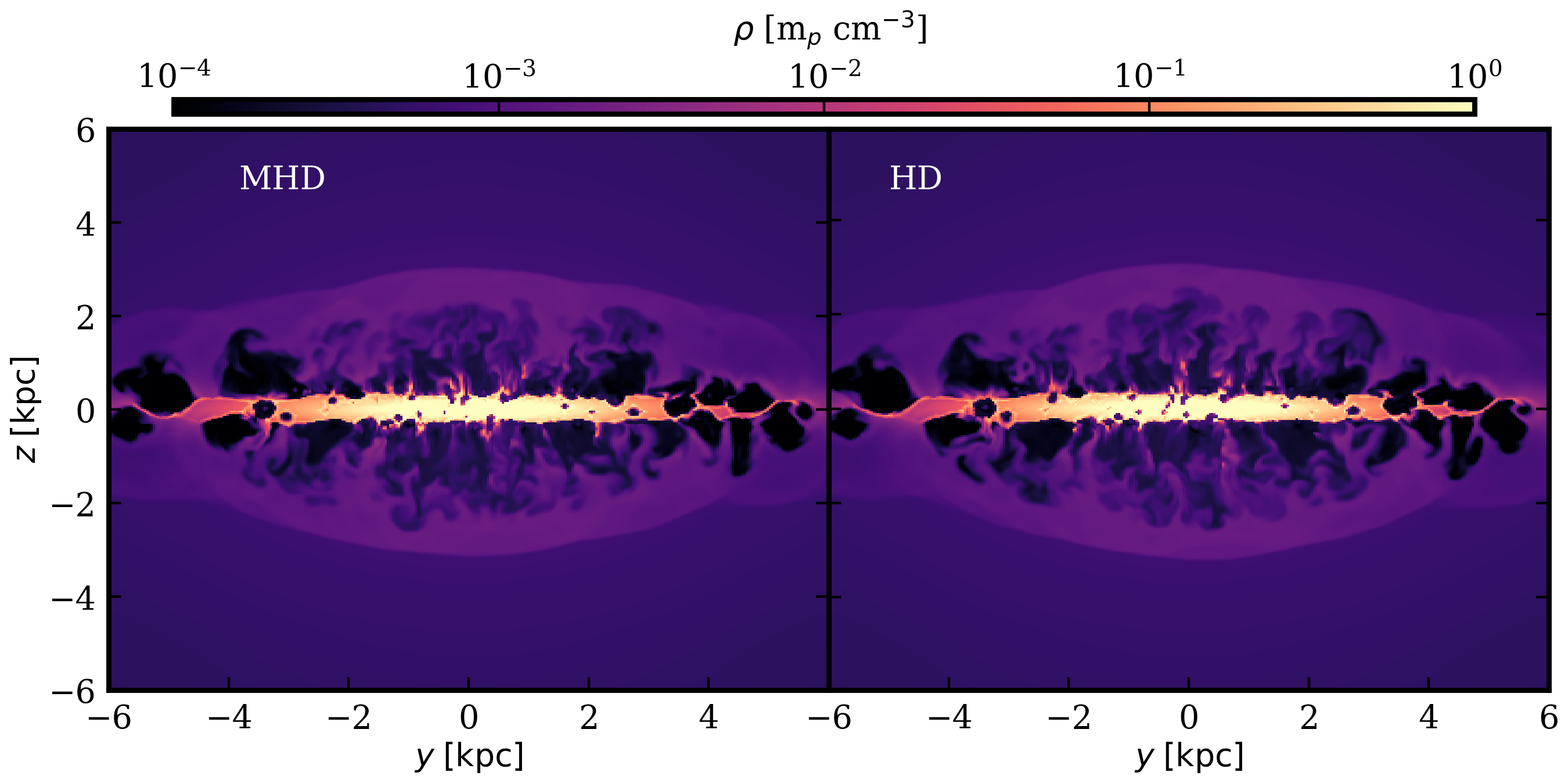}
	\caption{Density snapshots in the $x=0$ plane for MHD (left) and HD (right) runs at $5$Myr. Both the runs have identical initial conditions, resolutions and location of injection points. The SFR is $3$\sfr. Since the magnetic energy density is low these runs are very similar in morphology. }
	\label{fig:mhd_hd_dens}
\end{figure*}

In this paper, we address some issues regarding the morphology, vertical intensity profile of radio halos, such as its dependence on SFR surface density, for Milky Way type galaxies, with the help of magneto-hydrodynamical (MHD) simulations. Earlier simulations have looked at the dynamical and kinematical role cosmic rays (CRs) play in galaxy evolution \citep{Salem2014, 
Ruszkowski+17}. These simulations attempt to understand how CR pressure can affect large outflow structure. 
Our focus, however is on the role of galactic winds in the determining the structure of extra-planar magnetic field and the synchrotron emission profiles of edge-on star forming galaxies. The observations of \citealp{Dahlem1995} showed that the radial extent of the radio halo at some height above the disk remains the same as the radial extent along the disk. Therefore a disc-wide distribution of star formation sites is required to study the morphology. Our previous work on the effect of distributed star formation on the X-ray morphology \citep{Vijayan2018} provides us with a suitable set-up for this study, when magnetic field is added. 

\begin{figure*}
	\includegraphics[width=0.9\textwidth]{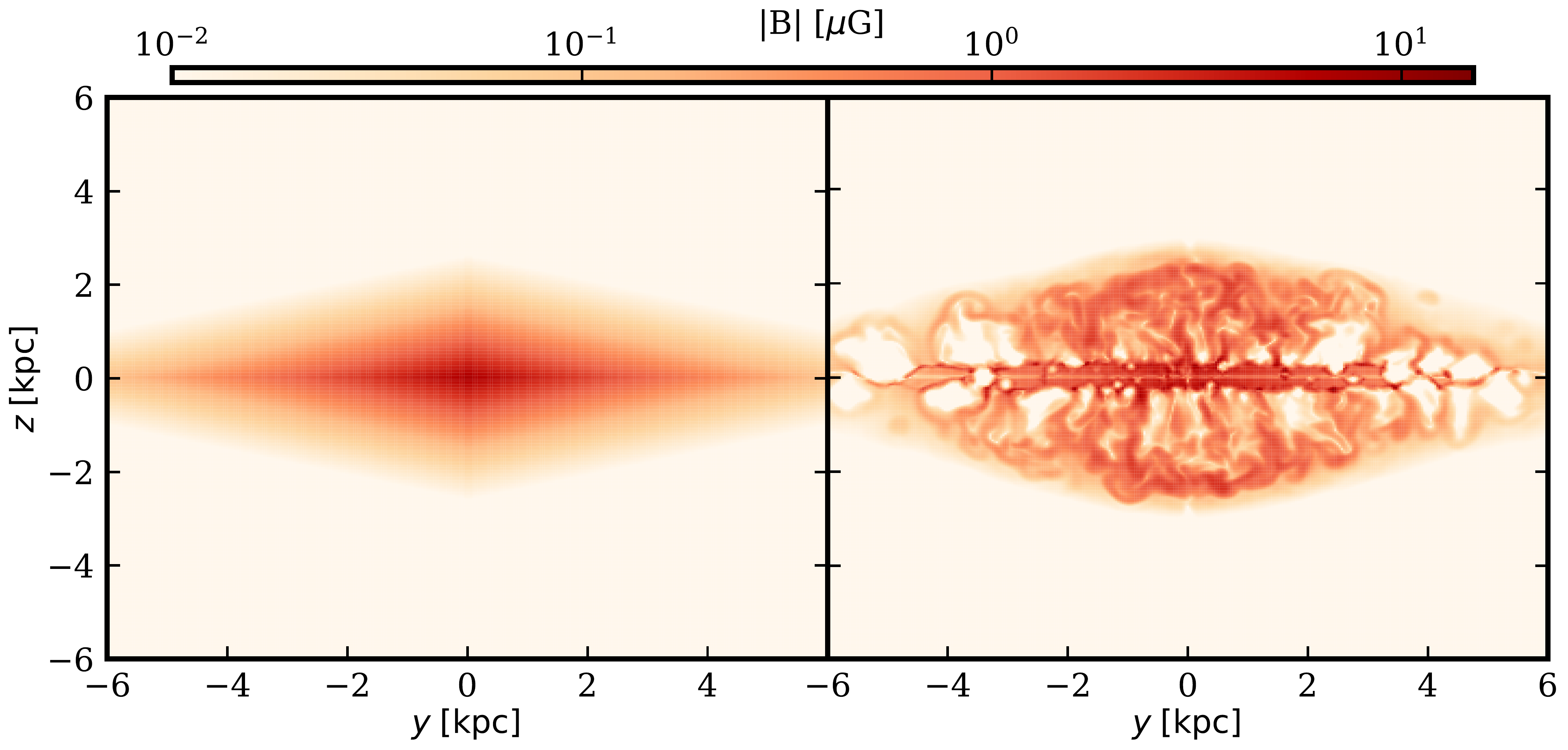}
	\caption{Magnetic field strength(|B|$=\sqrt{B_x^2 + B_y^2 + B_z^2}$) in $x=0$ plane at $t=0$ Myr (left) and at $t=5$ Myr (right) for the same SFR as in Figure \ref{fig:mhd_hd_dens}. The initial magnetic field has an exponential distribution (see Equation \ref{eqn:bfield}) and in the log scale field lines appear as straight lines with constant slope. The magnetic field in the disk gets advected with the gas resulting in magnetised outflows. Notice the unmagnetised bubbles ($y \sim 2-4$ kpc) due in injection of gas which does not contain magnetic field. Near the midplane, due to a number of injection points (see Section \ref{sec:num_setup}) a substantial region get depleted of magnetic field and that creates a low emissivity region close to $z \sim 0$.}  
	\label{fig:b_initial}
\end{figure*}

\section{Set-up}

\subsection{Disk and Halo}
The set-up is identical to the three dimensional set-up used in \citealp{Vijayan2018}. We simulate a Milky Way mass (M$_{\rm vir} = 10^{12}$ M$_{\odot}$) galaxy which comprises a high density, low temperature rotating stellar disk (M$_{\rm disk}= 10^{10}$ M$_{\odot}$) in equilibrium with a static halo. The halo gas is hotter ($T_{\rm vir}\sim 3\times10^{6}$ K) and less dense than the $4\times10^4$ K disk. The initial density and pressure profiles are determined by solving the equations of hydro-dynamic equilibrium between the rotating disk and the stationary halo. Mid-plane density is $3$ m$_p$ cm$^{-3}$. The stellar disk is modelled by Miyamoto-Nagai potential and NFW profile is used for the dark matter potential. For further details on the set-up, including the density profiles, the reader is directed to Section 2 of \citealp{Vijayan2018}.


\subsection{Initial Magnetic Field}
The magnetic field in the disk of our Galaxy is not very well understood. However, radio observations have been used to infer the probable topology and magnitude of the magnetic field. \citealp{Sun2008} describe a functional form of the Galactic magnetic field, in the disk as well as in the halo (their Equation 7), which we adopt here:
\begin{eqnarray}\label{eqn:bfield}
B_\phi&=&B_0 \exp \Bigl [-{R \over R_0} -{\vert z \vert \over z_0}\Bigr ] \,,\nonumber\\
B_z&=&B_R=0
\end{eqnarray}
Here $B_0$ is $2 \mu$G and $z_0$ and $R_0$ are equal to $400$ pc and $1.5$ kpc, respectively \citealp{Sun2008}. We have chosen the magnetic scale height, $z_0$, to be equal to the parameter $b$ of the disk potential. 
We have changed the configuration from \citealp{Sun2008}, in order to apply it to our simulated galaxy. While they had a constant magnetic field strength inside of a region $5.3$ kpc around the Galactic centre
and follow the exponential form outside (scaling as $\exp [-(R-R_\odot)/R_0]$), we have made it exponential from the central region. This is because in our simulated galaxy, the peak of star formation sites is at a galactocentric radius of $\sim 1$ kpc (while for our Galaxy, it is around $\sim 4$ kpc), and so the radial extent of our simulated galaxy is smaller than Milky Way.

The initial magnetic field has been set-up only in the disk region in using a vector potential. Since the thermal pressure is much higher than the magnetic pressure, the unbalanced Lorentz force leads to only a marginal ($<10\%$) change in the density distribution in the disk for the period of interest, $10$ Myr.  


\begin{figure*}
	\includegraphics[width=0.9\textwidth]{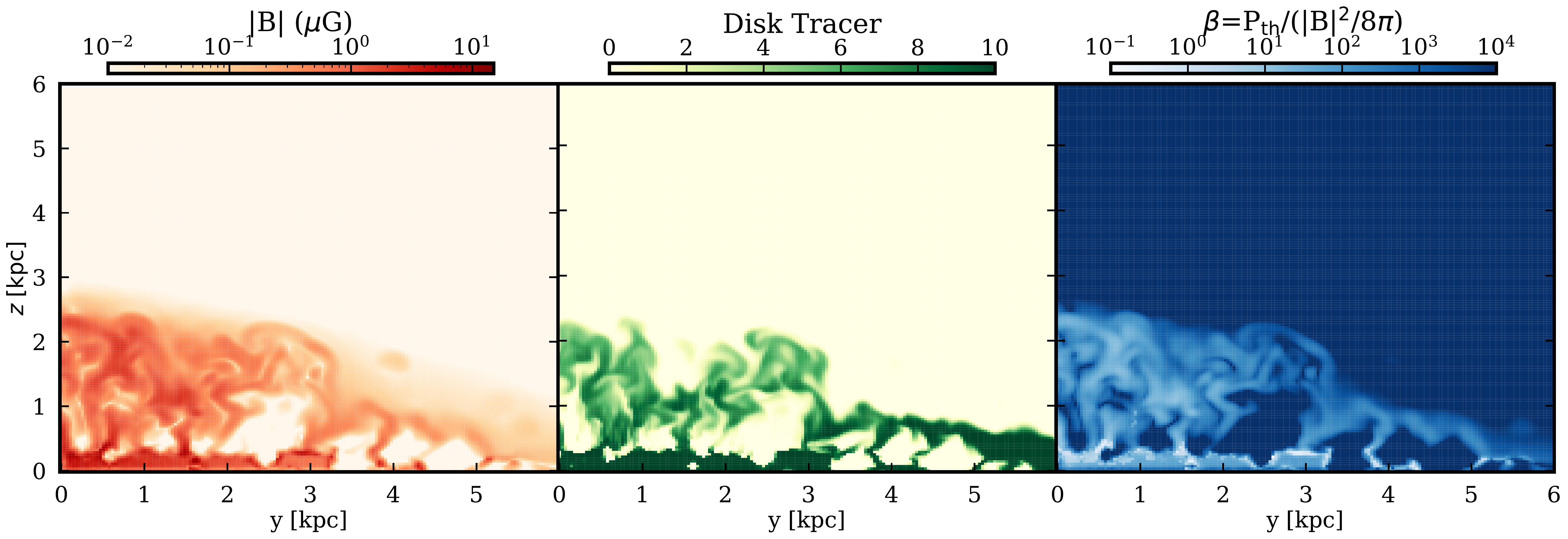}
	\caption{Magnetic field strength, $|B|=\sqrt{B_x^2 + B_y^2 + B_z^2}$, in $x=0$ plane at $t=10$ Myr (left) is compared with disk gas tracer (middle panel) in the same plane and the same time. Note that the field strength is high in regions where the disk tracer is high because magnetic field get lifted up along with the disk gas. Plasma $\beta$-parameter ($\beta =8\pi\rm{P}_{\rm th}/|{\rm B}|^2$), shown in the right panel, is lower in disk region, where magnetic field is high, and also in the regions where the magnetic field has been compressed along with the gas due to shocks.} 
	\label{fig:bxby_trc}
\end{figure*}

\subsection{Numerical Set-up} \label{sec:num_setup}

We use a $[-6,6]^3$ kpc$^3$ Cartesian grid which is uniform in all the three directions with a resolution of $36$ pc. We conduct all our simulations using the finite volume code, PLUTO \citep{Mignone2007}. We use HLLD (for MHD runs) and HLLC (for HD runs) solvers with linear reconstruction for all simulations discussed here. Constrained transport has been used to ensure $\vec{\nabla} \cdot \vec{B}=0$ throughout. Outflow boundary conditions are used everywhere.

The energy and mass injection rate have been chosen so that the total SFR across the disk is $0.3$ \sfr, $3$ \sfr, $30$ \sfr. We can estimate the mechanical energy and mass injection rate from the star formation rate by assuming Kroupa initial mass function and a factor $\sim 0.3$ of the total deposited energy to be thermalized \citep{Strickland2007, Vasiliev2019} as,
\be 
L_{\rm mech} \approx 10^{41} \, {{\rm erg} \over {\rm s}} \, \Bigl ( {{\rm SFR} \over {\rm M}_\odot \, /{\rm yr}} \Bigr ) \,.
\ee
Supernovae explosions and stellar winds also add mass to the surrounding interstellar medium, the precise amount of which is uncertain. We adopt a  mass injection rate that is related to the star formation rate in the following way,
\be
\qquad
\dot{M}_{\rm inj} =  0.1 \, {\rm SFR} \,.
\ee

We choose $1000$ injection points in the disk for mass and energy injection to simulate the locations of OB associations, and for simplicity, they have identical mass and energy injection rates. 
The location of these points have been chosen such that the Kennicutt-Schmidt law is followed  between the surface density of gas and surface density of star formation ($\Sigma_{\rm SFR} \propto \Sigma_{\rm gas}^{1.4}$). In the $z-$direction, the points have been distributed to follow the gas density. For an exponential gas distribution in the disk, this implies a radial distribution of injection points with a maximum at $\sim 1$ kpc for the parameters we have chosen (for details see Appendix A of \citealp{Vijayan2018}).

\section{Results}

We conduct 3-D simulations using the above described set-up. We compare the density snapshots for MHD and HD runs for our fiducial run with SFR$=3$ \sfr. In Figure \ref{fig:mhd_hd_dens} we show the density slices in the $x=0$ plane for MHD (left) and HD (right) runs after $5$ Myr of evolution. Star formation in each injection region creates a bubble of hot, low density gas around it. Over time bubbles from different injection regions merge and ultimately break out of the disk of the galaxy. The outflow generated is fastest close to the center of the galaxy because it has the highest density and a higher number of injection points. In the outer regions of the disk, the shock is not strong enough due to fewer injection points. Since the initial conditions-including distribution of injection points-is symmetric along the $z$ axis, the outflow is similar in $z>0$ and $z<0$. Morphologies of MHD and HD runs are very similar, which is expected since the magnetic energy density is smaller than the thermal energy density. As injected gas rises, it lifts up clumps of dense disk gas. These clumps can be clearly seen as patches of high density surrounded by a low density gas. As they move up, this dense disk gas gets mixed with the hotter injected gas.

\begin{figure*}
    \centering
    \includegraphics[width=\linewidth]{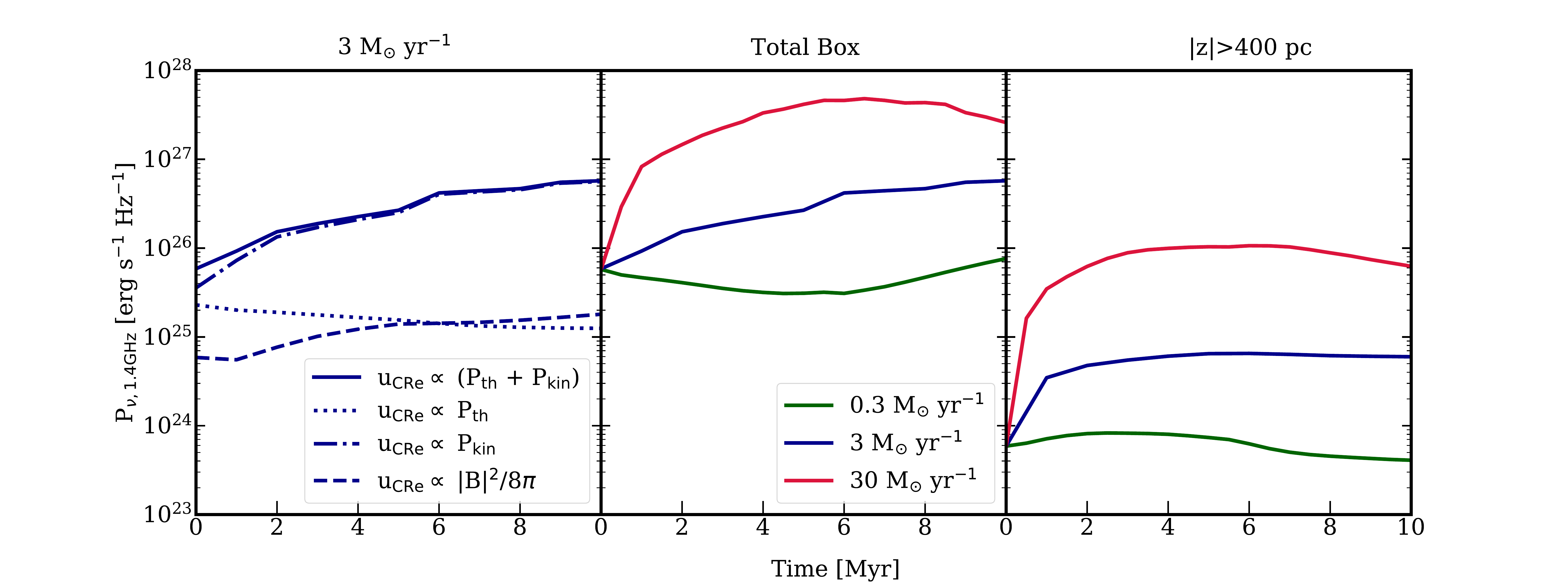}
	\caption{Evolution of total synchrotron power emitted at $1.4$ GHz for different expressions for \ucre (left) and SFRs (middle and left). The left panel shows the power emitted by the entire simulation domain for $3$ \sfr case by using \ucre proportional to thermal energy density (${\rm P}_{\rm th}$, dotted line), kinetic energy density (${\rm P}_{\rm kin}$, ${\rm P}_{\rm kin} = 0.5(v_x^2 + v_y^2 + v_z^2)$, dot-dashed line), total gas energy density (${\rm P}_{\rm kin} + {\rm P}_{\rm th}$, solid line) and magnetic energy density ($|{\rm B}|^2/8\pi$, dashed line) , (see Equation \ref{eqn:emissivity}). The total power estimated using the sum of thermal and kinetic energy gives a number closest to analytical estimates (see expression for total power, P$_{\nu}$, relation \ref{eqn:radio_lum_est}). Thus we use it to find the power for other SFRs - shown in the two right panels. The middle panels shows total power from the entire simulation domain for three SFRs - $0.3$ (green), $3$ (blue) and $30$ (red) curves. The red curves increases and then shows a decrease after  $\sim 5$ Myr once the outer shock moves out of the simulation domain. The green curve shows the exact opposite trend because of intense radiative losses in the disk before the outer shock can break out of the disk.
	The right panel shows the power emitted by the extra-planar region, which attains a steady state value within $2-3$ Myr. The green curve in this panel hardly changes from its $t=0$ value, thus exhibiting very low extended emission. 
	}
	\label{fig:totalL}
\end{figure*}

Magnetic field is entrained in disk gas as it traverses through the potential of galaxy. In Figure \ref{fig:b_initial} we show the magnitude of magnetic field ($|B|= \sqrt{B_x^2 + B_y^2 + B_z^2}$) in the $x=0$ plane (left panel) at $t=0$ (left panel) and at $t=5$ Myr (right panel) for the fiducial run.
We can see that the disk region still contains substantial magnetic field. However, in the regions above and below the disk its distribution is highly non-uniform. There are regions inside of the outer shock which are devoid of magnetic field (e.g., at $y \simeq 2-3$ kpc). The corresponding region in the density slice shows a drop in density. Thus, the magnetic field is preferentially high in regions of high density.

To further explore the spatial correlation between magnetic field and density, the left two panels of Figure \ref{fig:bxby_trc} show $|B|$ and the disk gas tracer for the first quadrant of the simulation in the $x=0$ plane. Tracers are scalar quantities that get advected with the fluid. We have used three kinds of tracers---for metallicity, disk gas and injected gas. Disk gas tracer has been set so that it has high value ($\sim 10$) in  cells which are dominated by disk gas at $t=0$ identified using a density criterion and low ($\sim 0$) otherwise. We find that in Figure \ref{fig:bxby_trc} the regions that correspond to high value of the disk tracer (green patches) have a high magnetic field (orange patches). This is expected because some of the disk gas is lifted up by outflows. 
The right panel of Figure \ref{fig:bxby_trc} shows the plasma-$\beta$ parameter, which is the ratio of the thermal pressure (${\rm P}_{\rm th}$) and magnetic energy density ($\beta = 8\pi {\rm P}_{\rm th}/|{\rm B}|^2$). Since the initial magnetic field is small, $\beta$ parameter is greater than unity everywhere, even close to the disk and behind the outer shock, where material gets accumulated. 

\subsection{Synchrotron Emission}
Cosmic Rays (CRs) are generated through supernova explosions in the disk of the galaxy. In the presence of magnetic field, cosmic ray electrons (CRes) emit synchrotron radiation. CR particles are believed to get accelerated at shocks and simulations have estimated that they can gain $\sim 10\%$ of the bulk energy density (\citealp{Caprioli2014}). Of this, CRes account for a fraction, $(m_e/m_p)^{(3-p)/2}$, depending on the slope of their energy spectrum, $p$ \citep{Persic2014}. For $p=2.2$, this fraction is $\sim 5\%$. Thus, given gas energy density, \ugas, $\epsilon_{\rm{CRe}}\equiv$\ucre$/$\ugas  is equal to $0.01\times 0.05=5\times 10^{-3}$. 

We can write down the emissivity, $j_{\nu}$, in terms of $\epsilon_{\rm{CRe}}$ and the magnitude of magnetic field(B) as (refer to equation 18.17, Longair, 1981)
\begin{equation}\label{eqn:emissivity}
\begin{split}
j_{\nu} (\rm{erg} \rm{s}^{-1} \rm{cm} ^{-3} \rm{Hz}^{-1}) = 1.7 \times 10^{-21} \times a(p) \times  k \times \Big(\epsilon_{\rm{CRe}}\times u_{\rm{gas}}\Big) \\
\times B^{(p+1)/2)} \Big(\frac{6.26\times 10^{18}}{\nu}\Big)^{\frac{(p-1)}{2}} \,.
\end{split}
\end{equation}
$k$ is the normalisation for the energy spectrum, i.e., \ucre$=\int_{m_e c^2}^\infty k E^{-(p-1)}dE$. $a(2.2) \sim 0.1$, from Table 18.1, Longair, 1981. We take $\nu=1.4$ GHz for the rest of our analysis. 

\subsubsection{Cosmic Ray Energy Density}\label{subsec:gasEden}
One has to necessarily resort to assumptions when calculating the synchrotron radio emission from MHD simulations. The objective is to find a proxy for CR energy density for which there are different recipes. We assume \ucre$\equiv \epsilon_{\rm{CRe}} \times$\ugas to be proportional to magnetic or gas thermal/kinetic energy density. Since kinetic energy is frame dependent, caution should be exercised while using it for getting \ucre. However, it can be a reasonable proxy if gas motions are turbulent, since internal shocks as well as adiabatic compression can give rise to cosmic rays. In this section, we discuss how these different prescriptions will affect the total radio luminosity.

\begin{figure*}
	\includegraphics[width=\textwidth]{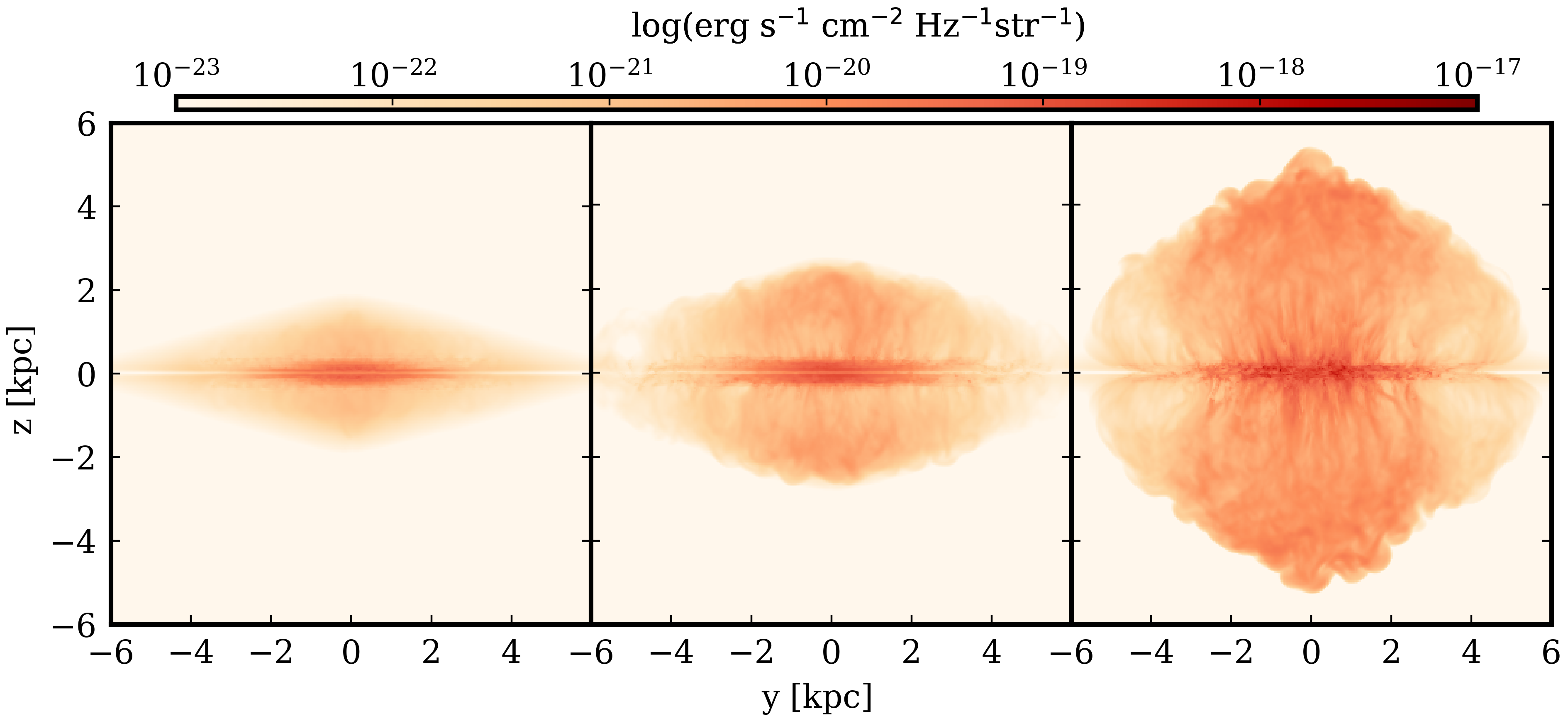}
	\caption{Surface brightness snapshots at $5$ Myr by taking \ugas$\propto ({\rm P}_{\rm th} + {\rm P}_{\rm{kin})}$ for SFR =$0.3$ \sfr, $3$ \sfr and $30$ \sfr (from left to right). There is hardly any extra-planar emission from $0.3$ \sfr case even after $5$ Myr (compare with left panel in Figure \ref{fig:b_initial}). Filamentary structure in emission is observed in $3$ and $30$ \sfr cases as channels are created by outflows from the disk. Enhancement of emission is behind the outer shock as a result of gas accumulation.}
	\label{fig:sb_dens}
\end{figure*}

To understand how these different prescriptions will affect the observables, in Figure \ref{fig:totalL} we show the radio power at $1.4$ GHz, P$_{\nu,1.4\rm{GHz}}$, for the simulation box as a function of time. The total power  is obtained by summing up the emissivity (given by Equation \ref{eqn:emissivity}) for each cell in the simulation domain and multiplying with the total volume of the box, P$_{\nu,1.4\rm{GHz}}=\int j_{\nu}$dV. We obtain different curves by taking \ucre  proportional to thermal pressure ($=1.5 {\rm P}_{\rm{th}}$), kinetic energy density ($=1.5 {\rm P}_{\rm{kin}}$, ${\rm P}_{\rm kin}=0.5 \rho({\rm v}_x^2+{\rm v}_y^2+{\rm v}_z^2)$ ) and sum of the two. In the left panel, we show total power for $3$ \sfr run from the entire box. The luminosity according to thermal energy density declines with time which is unphysical, since there is a continuous energy injection from star formation. Further, we note that the kinetic pressure component dominates the thermal pressure component of total power. 

In the dashed curve of the left panel of Figure \ref{fig:totalL}, we show the power if \ucre was in equipartition with magnetic field energy density ($=|{\rm B}|^2/8\pi$). The total power obtained by making this assumption falls short of analytical estimates (see expression \ref{eqn:radio_lum_est}). This is not surprising since the initial magnetic energy density is not in equipartition with the gas energy density.

The three curves in the right two panels of Figure \ref{fig:totalL}, show the temporal dependence of radio power for all the SFRs assuming \ucre$\propto ({\rm P}_{\rm th} + {\rm P}_{\rm kin})$. The middle panel  shows the power for the entire simulation domain and the right panel shows the contribution of the extra-planar gas ($|z|>400$ pc). Power for $30$ \sfr decreases after about $\sim 5$ Myr once the outer shock moves out of the box and outflows escape the simulation domain. For $3$ \sfr the power increases monotonically throughout the entire run of $10$ Myr. From Figures \ref{fig:mhd_hd_dens} and \ref{fig:bxby_trc}, we see that the disk contains substantial gas as well as magnetic field intensity even in the presence of outflows. We, therefore, expect the extra-planar component to be sub-dominant (right panel, Figure \ref{fig:totalL}). Once the outflows breach the scale height of the disk (after about $\sim 2$ Myr for $3$  and $30$ \sfr), the power emitted remains nearly a constant. In the following sections, we discuss how radio power depends on time as well as the magnetic field and SFR. 

In order to select an appropriate measure of \ucre, we can compare the corresponding radio power with one-zone model. We can estimate a conservative value of the synchrotron luminosity, by taking \ugas  equal to $L_{\rm mech} \times t$. P$_{\nu,1.4\rm{GHz}}$ can be written as,
\begin{equation}
\begin{split}
P_{\nu, 1.4 {\rm GHz}} \approx 4 \times 10^{26} \, {\rm erg} \, {\rm s}^{-1} \, {\rm Hz}^{-1}\, \Bigl({t \over 10 \, {\rm Myr}}\Bigr)\,\Bigl({L_{\rm mech} \over 3 \times 10^{41} \, {\rm erg} \, {\rm s}^{-1}}\Bigr )\,\\
\Bigl ({B \over \mu{\rm G}}\Bigr)^{1.6}\,.
\label{eqn:radio_lum_est}
\end{split}
\end{equation}
At $10$ Myr, the power estimated from using the sum of thermal and kinetic energy density gives $\approx 3\times 10^{26}\, {\rm erg} \, {\rm s}^{-1} $, while the estimated power is $ \approx 4\times 10^{26}$, for $B\simeq 1\mu$G.


 Comparison with the curves in Figure \ref{fig:totalL} (left panel) shows that  using only the thermal energy density  leads to an underestimate of the luminosity, and also produces a declining luminosity over time, contrary to the expectations from  Equation \ref{eqn:radio_lum_est}. 
 Using the sum of thermal and kinetic energy density provides a better match with the observed luminosity. The volume averaged magnetic field within the disk, dominant source of synchrotron emission, is a decreasing function of time. This is because the injected gas evacuates larger and larger un-magnetised holes. Therefore, the radio luminosity would not grow linearly in time, as seen from the green and blue curves in Figure \ref{fig:totalL}. 
 We also note that the estimate of the radio luminosity in Equation \ref{eqn:radio_lum_est} matches the observed values (e.g. refer to Table 3 of \citealp{Dahlem1995}). We need to scale the power by surface density of energy injection rate, $\dot{E_A}$, the radius of star formation, $r_{\rm SF}$, equal to $4$kpc for our simulations, and the magnetic field. We get $P_{\nu} \sim \dot{E_A} r_{\rm SF}^2 B^{1.6}$ and for NGC $4666$ ($r_{\rm SF}=13$ kpc, B$=8.8 \mu$ G) this comes out to be $\sim 5\times 10^{26}$ \ergps, comparable to the power at $10$ Myr for $3$\sfr. 

Therefore, we adopt the sum of kinetic and thermal energy densities as the gas energy density and use it for synchrotron emissivity for further analysis. We note here that the conclusion - that kinetic energy determines synchrotron power- follows from the fact that magnetic field energy density is much lower than the gas thermal energy density (see Figure \ref{fig:bxby_trc}, right panel). Observationally, it has been found that \ucre and magnetic field energy density are in equipartition \citep{Beck2012, Beck2005}.

\begin{figure}
	\includegraphics[width=\columnwidth]{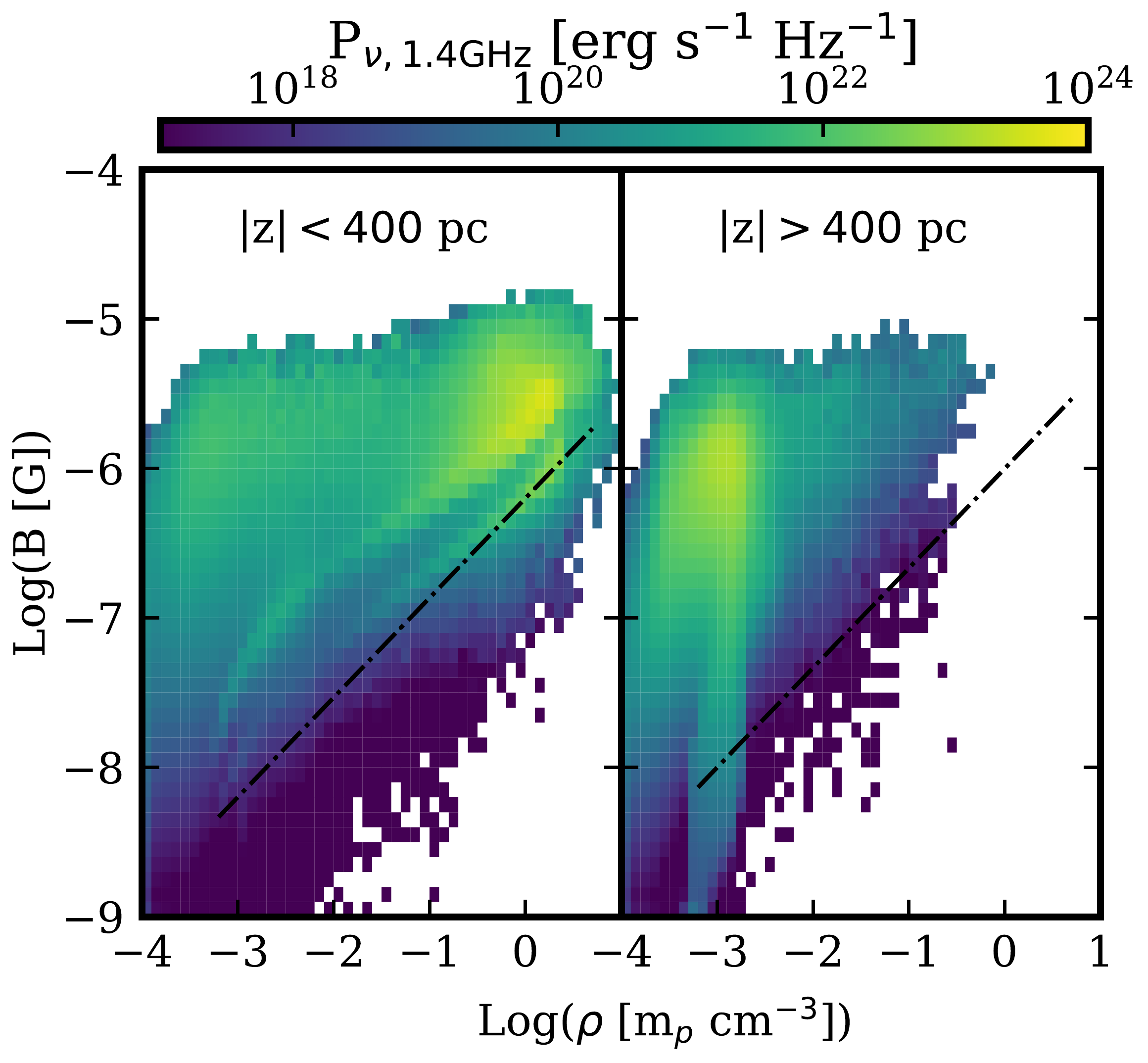}
	\caption{Scatter plot between density, $\rho$ (along horizontal axis) and magnitude of magnetic field, B, along vertical axis, colour-coded by synchrotron power for $3$ \sfr case at $5$ Myr. Left panel shows this distribution for the disk region ($z<400$ pc), while the right one is for the extra-planar region ($z>400$ pc). The dot-dash line represent power law relation between B and $\rho$, B$\propto \rho^{2/3}$ (assuming isotropicity), expected from flux freezing. The gas in the disk follows this relation due to flux freezing. The magnetic field strength of the extra-planar gas has been diluted by the injected gas and does not follow this relation very strongly.}
	
	\label{fig:b_rho}
\end{figure}

\begin{figure}
    \centering
	\includegraphics[width=\columnwidth]{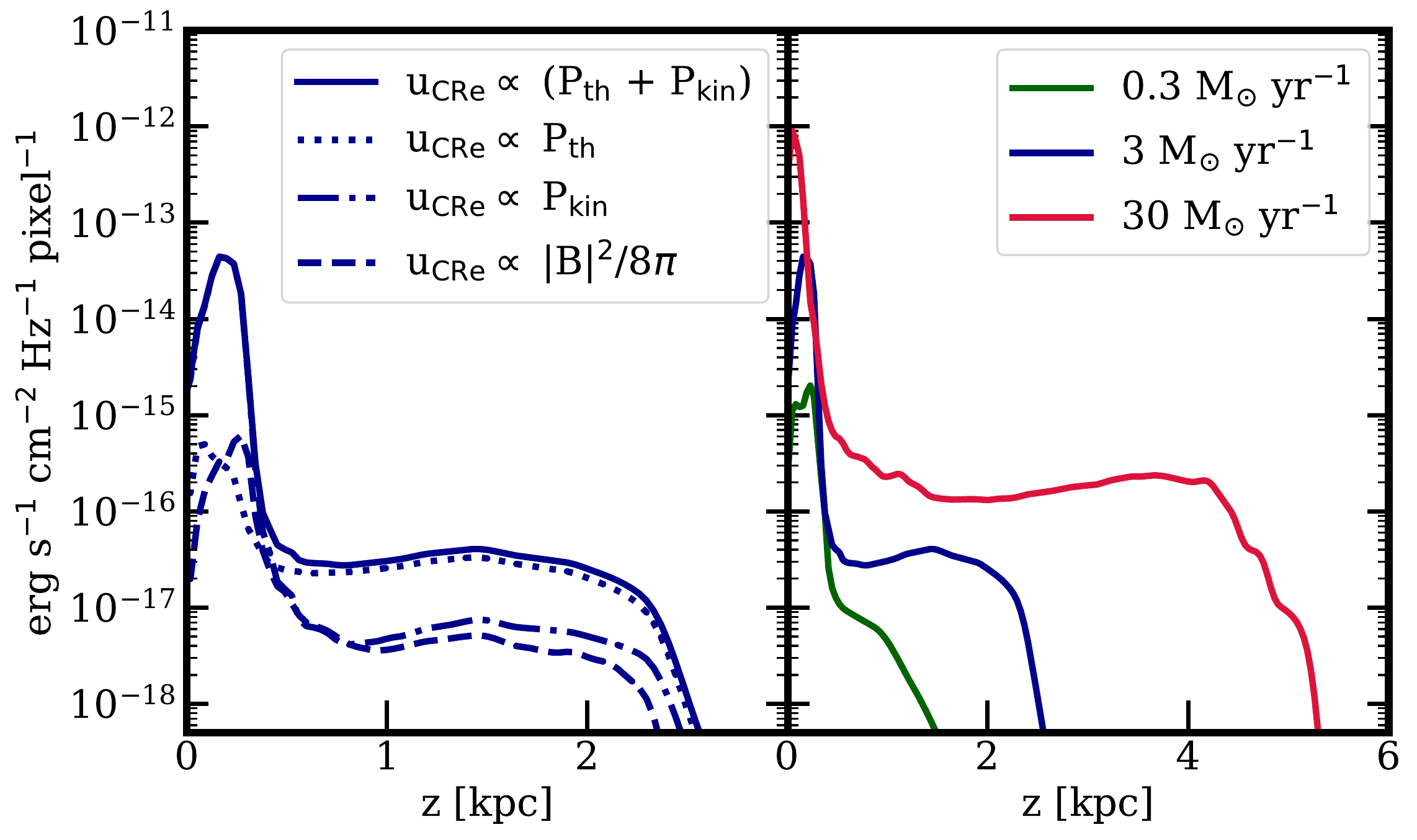}
	
	\caption{Vertical intensity profiles for SFR = $0.3$ \sfr, $3$ \sfr and $30$ \sfr at $5$ Myr. In the left panel, the dotted, dot-dashed, dashed and solid curves have been obtained in the same way as for Figure \ref{fig:totalL}. We see that near the mid-plane the emission is dominated by kinetic energy, while thermal energy is responsible in the higher $z$ regions. The right panel shows the vertical intensity for the three SFRs - $0.3$ , $3$ and $30$ \sfr - in green, blue and red curves. The red and blue curves show a characteristic feature - plateau followed by a slight bump. }
	\label{fig:zbrit}
\end{figure}

\begin{figure*}
    \includegraphics[scale=0.35]{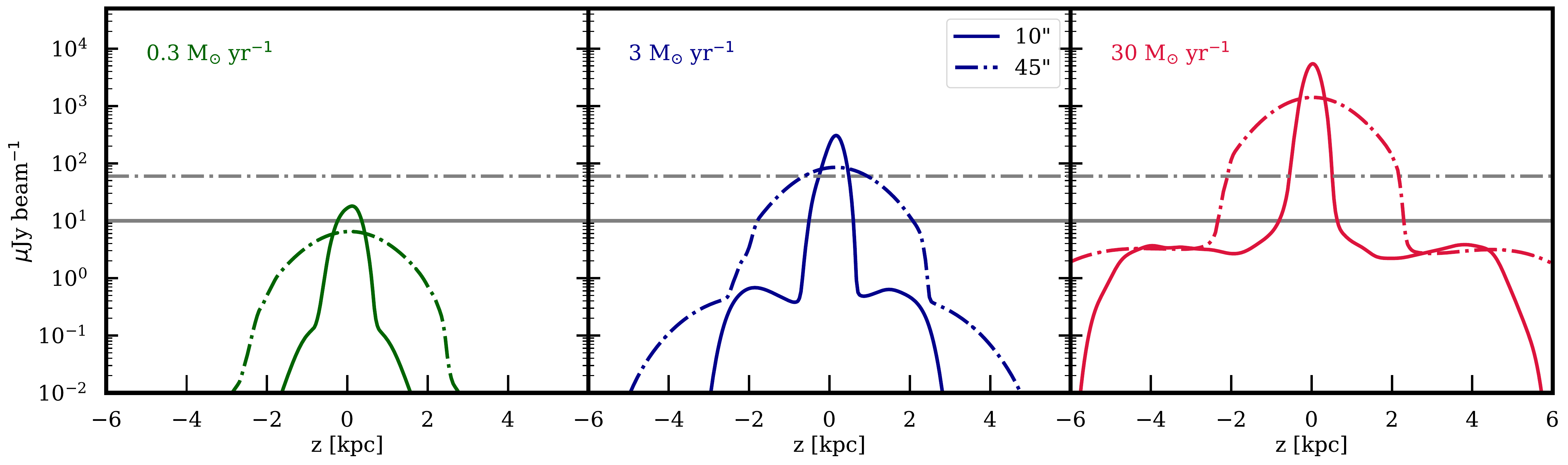}
	
	\caption{ Vertical intensity profiles for different SFRs and different choices for \ugas (see Section \ref{subsec:gasEden}) $0.3$ \sfr(left), $3$ \sfr(middle) and $30$ \sfr(right) convolved with beams of FWHM $10$" (solid) and $45$" (dot-dahsed). The horizontal lines show the RMS noise levels at $10 \mu$ Jy (solid) and $60 \mu$Jy (dot-dashed). We assume that the galaxy is at a distance of $10$ Mpc from Milky Way. The plateau feature is undetectable even for the highest star formation we have used. 
	}
	\label{fig:convol}
\end{figure*}

Using Equation \ref{eqn:emissivity} for emissivity, we can produce surface brightness maps for the outflows. Surface brightness is obtained by multiplying emissivity and the elemental length along the line of sight, taken to be $x$-axis. In Figure \ref{fig:sb_dens} we show the surface brightness at $5$ Myr for for different SFRs, increasing from left to right. Due to the presence of strong magnetic field, the disk is the strongest emitter of synchrotron radiation. The emission decreases in regions above and below the disk as the injected gas pushes on the disk gas and fills the volume within the outer shock. At much higher heights, $z\sim 4-5$ kpc, the material collects in the turbulent region behind the outer shock. There is insignificant extra-planar emission for the lowest SFR but its extent increases with SFR.
This results in increased emission at this height, which we refer to as the 'bump'. The observational consequences of the bump are discussed in Section \ref{sec:scale_H}. One can compare the simulated surface brightness map in Figure \ref{fig:sb_dens} with the high resolution images of edge-on starburst galaxies. For example that of NGC $4666$ (see Figure 6 in \citealp{Irwin1999}), which shows filamentary structure because of chimneys created during the break-out of gaseous outflow from the disk. Such a filamentary structure, extending outwards from the disk, is visible in two right panels of Figure \ref{fig:sb_dens}. 

In Figure \ref{fig:b_rho} we show the scatter plot between the density and magnitude of magnetic field for $3$ \sfr case at $5$ Myr. The plot has been colour-coded by synchrotron power. We separate the simulation domain into disk (right panel) and extra-planar region (left panel). The dot-dashed lines in both the plot have a slope of $\rho^{2/3}$. The distribution of high emissivity cells in the disk region follows closely the $\rho^{2/3}$ slope. In the extra-planar regions, we see adherence to the relation- especially as extension in the high density region of the plot. It should be noted that in the extra-planar region, the disk gas has been mixed with the fresh injected gas which carries no magnetic field. We therefore do not expect the relation, arising out of flux freezing, to apply very strongly in $z>400$ pc region.

\subsection{Scale Height of Emission}\label{sec:scale_H}

Figure \ref{fig:zbrit} shows the vertical intensity profile of radio emission. We obtain vertical intensity profiles by summing the emissivities in the $x-y$ plane in a slab of thickness $36$ pc and then plotting it along the $z$ axis. The curves in the left panel show this profile for SFR of $3$ \sfr at $5$ Myr using different prescriptions for \ucre.  The thermal energy density in the disk is low due to presence of cold gas and the kinetic energy of the rotating disk is expected to dominate. However, as the gas is lifted up to extra-planar height, we see the reverse. The right panel shows the profile for the different star formation rates. As visible in surface brightness plots, the extent of emission increases with SFR. The bump feature mentioned in the previous section is clearly seen in vertical profiles, especially in the red curve.

In order to  compare these simulated vertical intensity profile images with observed images, one has to convolve the simulated images with typical beam sizes of radio telescopes. The vertical intensity profile $I_\nu (z)$ is convolved with a Gaussian of width $\sigma$ to yield the convolved profile:
\begin{equation}
    I_{\nu, \rm convolved} (z_1)={\int I_\nu (z) \,{ \exp \Bigl [ -{(z-z_1)^2 \over 2\sigma^2} \Bigr] \, dz} \over {\int \exp \Bigl [ -{(z-z_1)^2\over 2 \sigma^2}\Bigr ] dz} }  \,,
\end{equation}
where $\sigma={\rm FWHM}/2\sqrt{2}$, and the full-width half maximum (FWHM) is obtained by the length subtended by the angle of the beam at the distance of the galaxy.
For this purpose, we take the example of a well-known and nearby edge-on star forming galaxy, NGC $891$, which is also a Milky Way type galaxy (rotational speed of $\sim 220$ km s$^{-1}$ ), located at a distance of $9.5$ Mpc \citep{Fraternali2011}. We assume a distance of $10$ Mpc. For a typical radio telescope, we use the parameters of the Giant Meter-Wave Radio Telescope (GMRT), which has a synthesized beam of $10''$ ($45''$) at L-band, with a RMS noise per beam of $\sim 10$($60$) $\mu$Jy for the entire (compact) array (private communication with K. S. Dwarakanath). Figure \ref{fig:convol} shows convolved vertical intensity profiles from Figure \ref{fig:zbrit} for beams with FWHM of $10''$ (solid) and $45''$ (dot-dashed). The corresponding horizontal  lines show the RMS noise per beam for GMRT.

The extra-planar emission is unlikely to be detected with either the synthesized beam or with the beam produced by the compact array of GMRT. The emission becomes prominent with increasing SFR. However, even for a high SFR, the extra-planar radiation is below the noise level, making it difficult to detect. It is then easy to understand why the observed profiles never reported such an enhanced patch of emission above the gas scale height. It is likely to be detected for very high SFR, for starburst galaxies. The nearest galaxy with a SFR of $\sim 10$ \sfr, M82, has never been imaged with attention given to extended low surface brightness features. It will be an interesting test of our result if a broad beam is used in order to detect such features. Such an accumulation of matter has been observed in other wavebands. M82's vertical intensity profile for CO intensity also shows a bump above the gaseous scale height (see Figure 11 of \citealp{Leroy2016}), similar to the continuum emission plateau plotted in Figure \ref{fig:convol}.

It is interesting to note that NGC $4631$, an edge-on galaxy 
has been observed with the Ooty Synthesis Radio Telescope, and its vertical intensity profile shows a plateau of the type we have described above \citep{Sukumar1985}. They did not observe the plateau at $1.4$ GHz, possibly because of lower flux density at high frequency. It will be interesting to observe this galaxy again with better (and more sensitive) instruments. Since current observations cannot detect the low surface brightness plateau, we use convolved profile in order to estimate the radio scale height. 

\begin{figure}
\centering
	\includegraphics[scale=0.4]{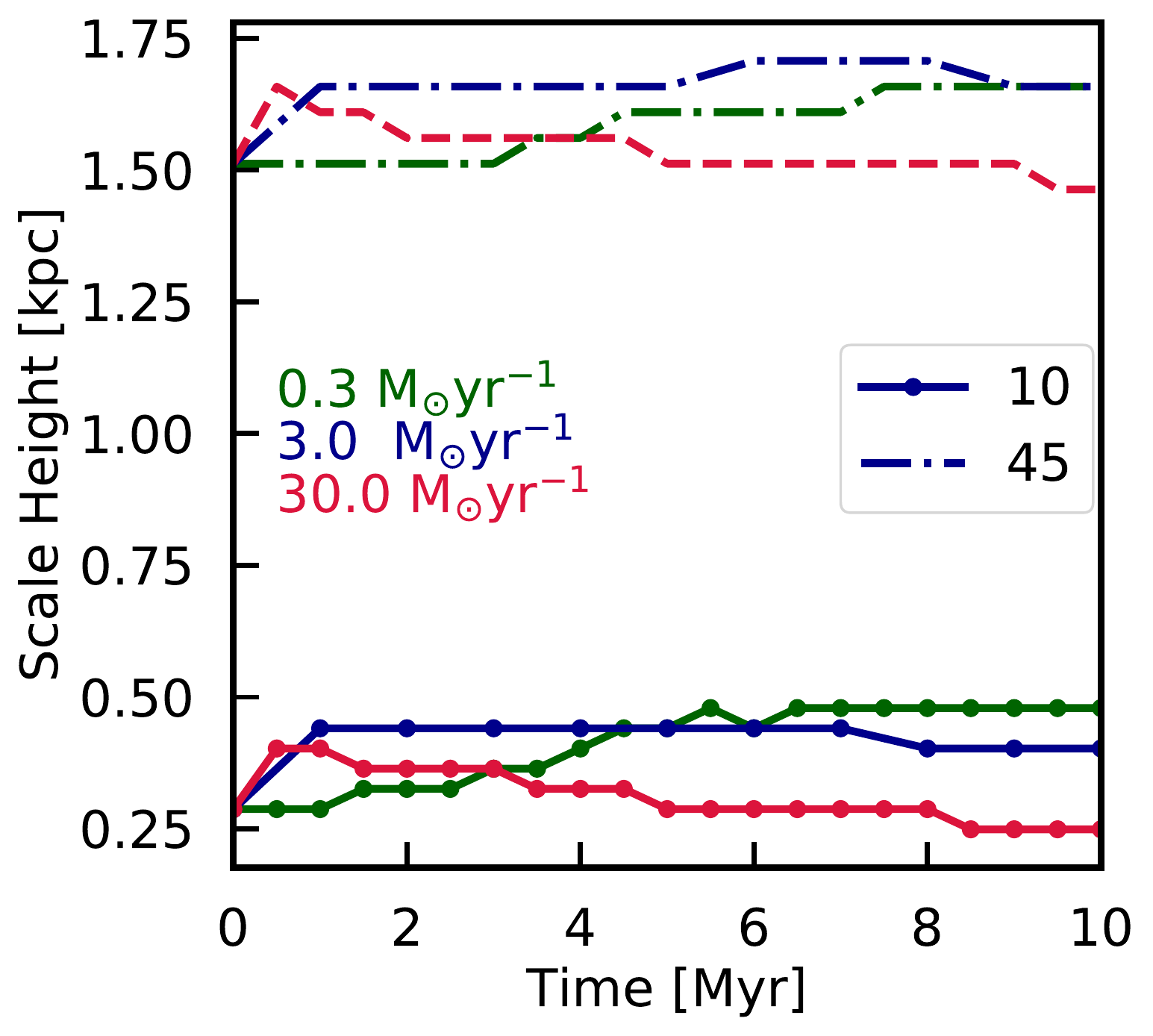}
	
	\caption{Variation of scale heights with time for $0.3$, $3$ and $30$ \sfr (red and blue, respectively). The dashed lines for a resolution of $45$" and dots represent a resolution of $10$". Scale height has been found using Figure \ref{fig:zbrit} (see Section \ref{sec:scale_H}). We see it reflects the beam size.
	}
	\label{fig:scale_H}
\end{figure}

In Figure \ref{fig:scale_H} we show how the radio emission scale height changes with time for different SFRs and beam sizes. We have determined the scale height as the height at which the convolved vertical intensity profiles of Figure \ref{fig:zbrit}, fall by a factor of $e$. For a higher resolution (FWHM $\sim 10''$), the scale height is between $250 - 500$ pc for all the three star formation rates, while for a larger beam size it lies between $1.5 - 1.75$ kpc.  The scale heights remain close to the linear resolution represented by beam sizes ($0.5$ and $2.2$ kpc on an object at a distance of $10$ Mpc for $10$" and $45$" respectively). The slight variation in scale heights with time is a result of changing nature of outflows.
 
\begin{figure}
    \centering
	\includegraphics[scale=0.4]{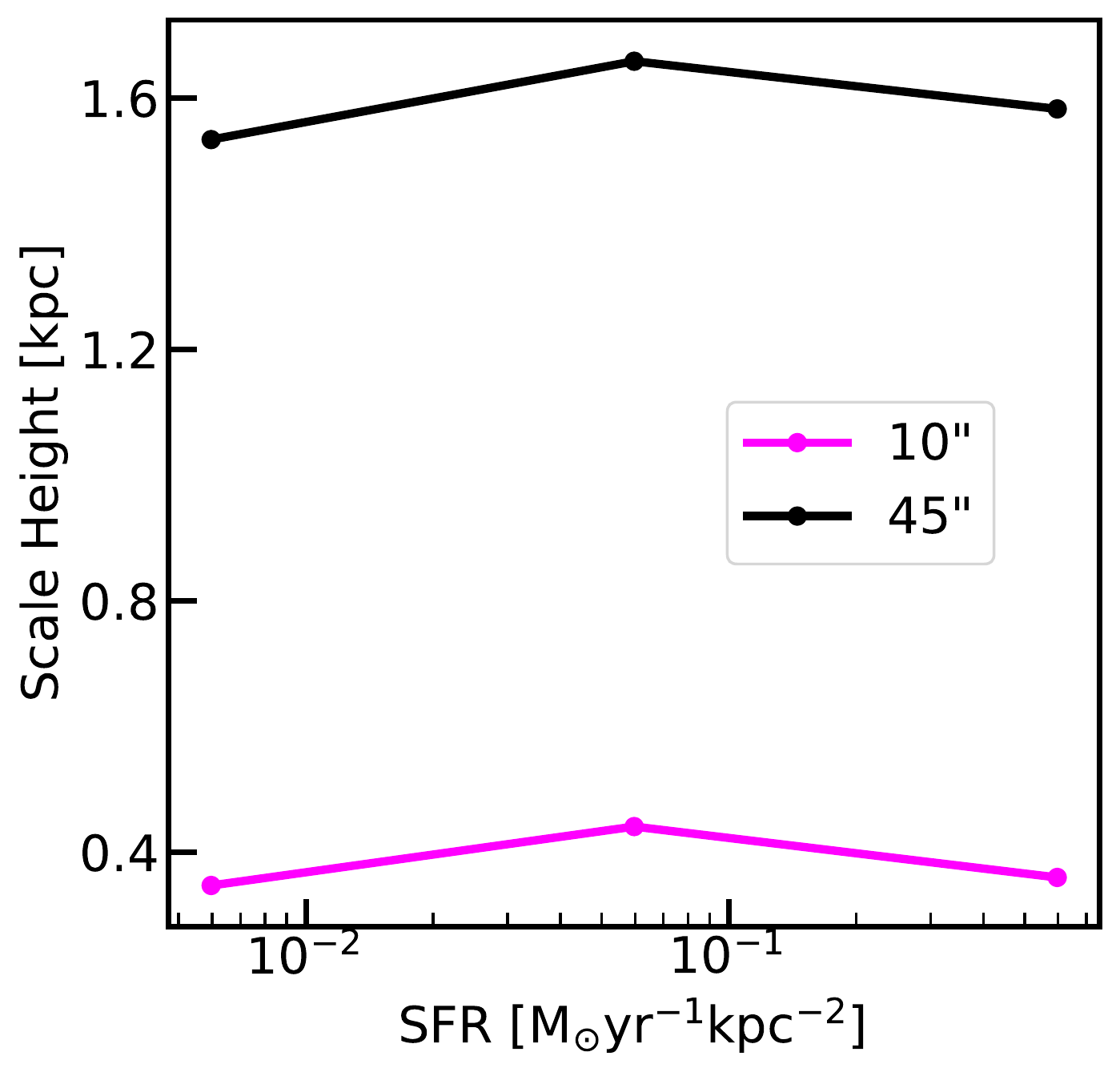}
	\caption{Variation of scale heights with SFR for two different resolutions - $10$" and $45$". The scale height is independent of the SFR.
	}
	\label{fig:scaleH_sfr}
\end{figure}

In Figure \ref{fig:scaleH_sfr} we show the time-averaged radio emission scale height for three star formation rates.  \citealp{Krause2018} presented detailed results for NGC $5775$, which has the same morphological type as MW (although with a much higher SFR $\sim 90$ \sfr) at a distance of $20$ Mpc, in the L-band of VLA. They use a resolution of $10$" and report the L-band scale height as $\sim 2$ kpc. To make a comparison with Figure \ref{fig:scale_H}, we would require an equivalent resolution of $20$", with our fiducial galaxy at $10$ Mpc. It is therefore reasonable to conclude that the scale height would lie between the magenta and black curves in Figure \ref{fig:scale_H}, and is roughly consistent with the reported scale height.

\subsection{Advection Speed}

\begin{figure}	
    \centering
	\includegraphics[scale=0.4]{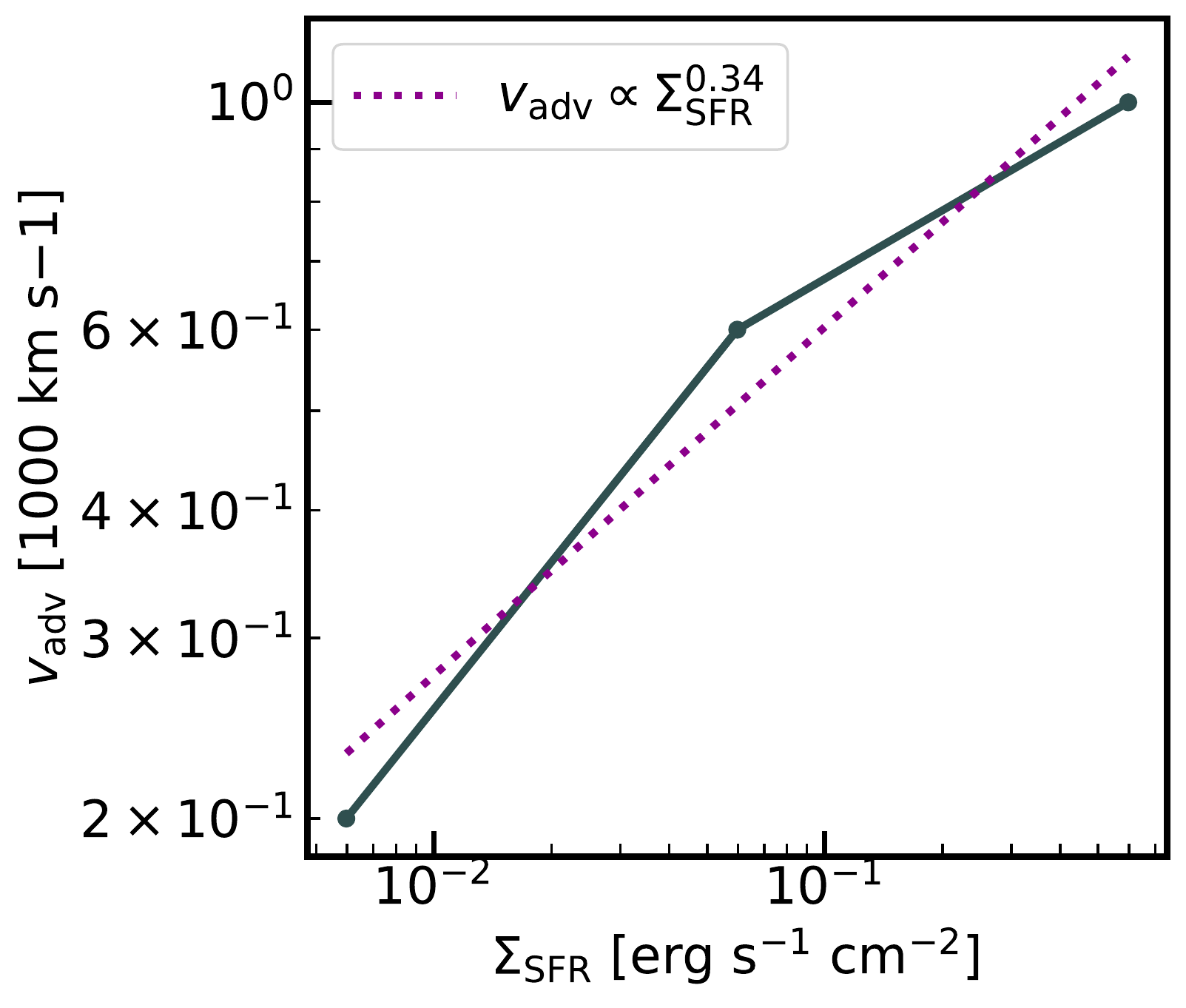}
	\caption{ Advection speed ($v_{\rm adv}$) versus SFR surface density is shown by the blue curve. Advection speed is measured by estimating the distance travelled by the outer shock in $5$ Myr. The dotted line shows a power law fit with $v_{\rm adv} \propto \Sigma_{\rm SF}^{1/3}$. }
	\label{fig:speedsfr}
\end{figure}

We show the variation of advection speed of the outer shock with SFR surface density in Figure \ref{fig:speedsfr}. 
The speed is estimated by the distance travelled by the outer shock in $5$ Myr. The curve shows that the speed scales with SFR surface density roughly as ${\rm v}_{\rm adv} \propto \Sigma_{\rm SFR}^{1/3}$. This is consistent with the observations of \citealp{Heesen2016}, and can be understood from a similarity analysis of a planar blast wave \citep{Vijayan2018}. Consider the self-similar motion of a one-dimensional shock, driven by energy injection in the disk through the extra-planar region. The relevant parameters for the evolution of the shock are the halo density $\rho$, the SFR surface density $\Sigma_{\rm SFR}$ and time $t$. These parameters can be combined to give the height of the shock at a given time as, $R \propto (\Sigma_{\rm SFR} t^3/\rho)^{1/3}$, which gives the speed of the shock as ${\rm v}_{\rm adv} \propto \Sigma_{\rm SFR}^{1/3}$, as observed, and as found in our simulation.

Advection of gas into the extra-planar region in the above mentioned manner has an interesting implication. Once the disturbance due to injection of matter and energy in the disk has been able to break out of the disc, the energy that is being deposited due to star formation is readily transported to the extra-planar region. Consider again a parallel slab geometry, and the time evolution of the gas in the extra-planar region. The above considerations show that the height scales with time as $R\propto t$, which implies (since the radial extent of the extra-planar region does not change with time) that the volume of the extra-planar gas also scales as $V \propto t$. The total energy content of the gas increases linearly with time, since the SFR is assumed to be constant. This means that the gas pressure in the exgtra-planar region stays roughly constant with time (since volume and total energy both increase linearly with time). Also, the energy content of the CRes increases linearly with time. However, due to flux freezing, the magnetic field in the extra-planar region decreases with volume as $B \propto V^{-2/3}$ and therefore, with time as, $B \propto t^{-2/3}$. The synchrotron luminosity of the extra-planar gas depends on the total energy content of the CR electrons (which scales as $t$) multiplied by $B^{1.6}$ (which scales roughly as
$t^{-1}$, because of the above mentioned scaling. These two factors cancel to give a roughly constant luminosity for the extra-planar gas with time (see Figure \ref{fig:totalL}).

\section{Discussion}

Our conclusions for the radio scale height have important implications for the shape of the radio halo of star forming galaxies. Since diffusion time scale is much longer than the advection time scale on $>$ kpc scales, the radial and vertical extents of the radio halo are unlikely to be different from as represented in our paper, even in the presence of diffusion. The radial extent of the radio halo is, therefore, determined by the extent of the star forming regions in the disc, {\it i.e.}, $4$ kpc in our simulation. Therefore a radio scale height of $0.4\hbox{--}2$ kpc implies a ratio of major to minor axis of $\sim 2\hbox{--}10$. This is consistent with the findings of \citealp{Singal2015} of major to minor ratio in edge-on star forming galaxies of being $\approx 2.5\pm 1.1$, also similar to a ``flattened ellipsoid'' inferred by \citealp{Weigert2015} from stacking 30 edge-on spiral galaxies. The near spherical-halo model of Milky Way of \citealp{Subrahmanyan2013}, proposed to explain the radio excess of \citealp{Kogut2011} without invoking any new population of sources or a cosmological origin, requires, in contrast, a ratio of $1.24$. Therefore our findings point out a discrepancy with the almost-spherical halo model of star forming galaxies and create issues for non-cosmological origin of the radio excess. If we consider diffusion, to make the radio halo more spherical (making the major to minor ratio $1.24 \pm 0.09$), the diffusion coefficient would have to be $\gtrsim 10^{29}$ cm$^{2}$ s$^{-1}$, even for an unrealistic long duration of star formation of $50$ Myr.

Our results are in agreement with the star formation rate surface density put forward by \citealp{Dahlem1995} for existence of radio halo ($\Sigma_{\rm SFR} \gtrsim 10^{-2}$ erg s$^{-1}$ cm$^{-2}$). For our lowest SFR case ($0.3$ \sfr ) $\Sigma_{\rm SFR}$ is equal to $6\times 10^{-3}$ \sfr kpc$^2$, since our star formation sites are confined within $4$ kpc. The total extra-planar emission from this case is nearly constant with time (right panel, Figure \ref{fig:totalL}) and increases as for other cases. We conclude that despite $10$ Myr of energy injection, a radio halo could not be sustained for this star formation rate surface density. 
However, this threshold is not independent of galaxy mass. It depends on two velocity scales: the advection speed, which scale as $\Sigma_{\rm SFR}^{1/3}$ (discussed above) and the escape speed, ${\rm v}_{\rm escape} \propto M^{1/3}$, with the galaxy mass $M$. Thus, the threshold energy injection surface density for the existence of radio halo should be smaller for a lower mass galaxy.

A caveat of our simulation is that magnetic field is not in equipartition with the turbulent and thermal energy density, initially. Further, we are not injecting any magnetic fields even though supernovae ejecta is expected to be magnetised. Therefore, our simulation is inadequate for addressing radio emission from the halo at larger scale heights as observed by \citealp{Hodges-Kluck2018} in the case of NGC $891$.  

\section{Summary \& Conclusions}
We conduct 3D MHD simulations of an isolated Milky Way type galaxy for three different star formation rate, viz, $0.3$, $3$ and $30$ \sfr. Our initially rotating gas disk comprises a smooth magnetic field (given by Equation \ref{eqn:bfield}) in the azimuthal direction. Upon comparison with hydrodynamic simulations with identical initial conditions, we see that morphology of outflows are not altered significantly due to magnetic field (Figure \ref{fig:mhd_hd_dens}). However, the outflowing gas carries magnetic field (see Figure \ref{fig:b_initial}). 

Such magnetised outflows lead to presence of radio halos around star forming galaxies as a result of synchrotron emission from cosmic ray electrons under the influence of magnetic fields. 
By assuming that cosmic ray energy density (\ucre) is a constant fraction of gas energy density (\ugas), we can estimate synchroton emissivity (using Equation \ref{eqn:emissivity}) for all cells in the simulation domain. We explore how different definitions of \ugas, such as being equal to thermal pressure (P$_{\rm th}$), kinetic pressure (P$_{\rm kin}$) or sum of both, can affect the macroscopic observables such as total power emitted (Figure \ref{fig:totalL}) and vertical intensity profiles (Figure \ref{fig:zbrit}). We conclude that using \ugas proportional to the sum of P$_{\rm th}$ and P$_{\rm kin}$ can match power with current observations. 

Our vertical intensity profiles show a characteristic feature which is a consequence of accumulation of material behind the outer shock. The profiles for all the star formation rate show a plateau (see e.g., red curve Figure \ref{fig:zbrit} in the vertical range $z\sim 2-4$ kpc for $30$ \sfr), and then a slight bump. Such a feature has not been reported in radio observations.

In order to compare our simulation results with the existing data, we produce these profile convolved with appropriate beam size ($10$" and $45$"). The convolved profiles are shown in Figure \ref{fig:convol}. While the convolved profiles faithfully reproduce the plateau plus bump feature, the RMS noise levels of instruments (compact array of GMRT) and are unlikely to be detected. This issue can be circumvented for a nearby starburst galaxy, such as NGC $4631$. In future, we aim to make observations for this galaxy to test our prediction.

Synchrotron emission has been studied, previously, by measuring the scale height of the radio halo. We use convolved profiles to understand how scale heights vary in time (Figure \ref{fig:scale_H}) and with SFR (Figure \ref{fig:scaleH_sfr}). The scale heights depend on the beam size, because the profiles are under-resolved. 

The main conclusions of our paper are as follows-
\begin{enumerate}
    \item Radio halos are formed as a result of magnetised outflows due to star formation in the disk of a galaxy. Our results of radio halo formed in Milky Way sized are broadly consistent with lower limit of energy injection surface rate density of $\gtrsim 10^{-2}$ \sfr kpc$^{-2}$ as was observed by \citealp{Dahlem1995}. We argue that this threshold scales linearly with galaxy mass, explaining the existence of radio halo in smaller galaxies with lower surface star formation rate densities (see \citealp{Heesen2016}). 
    
    \item We see an accumulation of magnetised gas behind the outer shock, resulting in enhanced synchrotron emission from higher $z$ regions of the galaxy. We show that this feature is difficult to observe because of its low surface brightness and extended nature and may have been missed in earlier observation. A notable exception is NGC $4631$, which should be revisted in order to test our prediction.
    
    \item We find that the scale height of radio halo is independent of SFR, which is consistent with observations. 
    
    \item We find that the major axis to minor axis ratio is $\sim 2-10$ which points to existence of an oblate radio halo for a star forming Milky Way mass galaxy. Spherical halos around the Galaxy have been cited to explain radio excess observed at the Galactic poles. 
    
    \item The advection speed of the radio emitting outflowing gas scaled with SFR as ${\rm v}_{\rm adv}\propto \Sigma_{\rm SFR}^{1/3}$. Such a scaling relation has been observed previously (e.g., \citealp{Heesen2018}).

\end{enumerate}

\section*{Acknowledgements}

We would like to thank Bhargav Vaidya and Dwarakanath K. S. for suggestions and inputs. We would also like to thank the anonymous referee for inputs and suggestions. Some of the simulations discussed in this paper were conducted on the SahasraT cluster at the Super-computer Education and Research Centre, Indian Institute of Science.

\appendix
\section{Convergence Test}

\begin{figure*}
    \centering
    \includegraphics[width=0.9\linewidth]{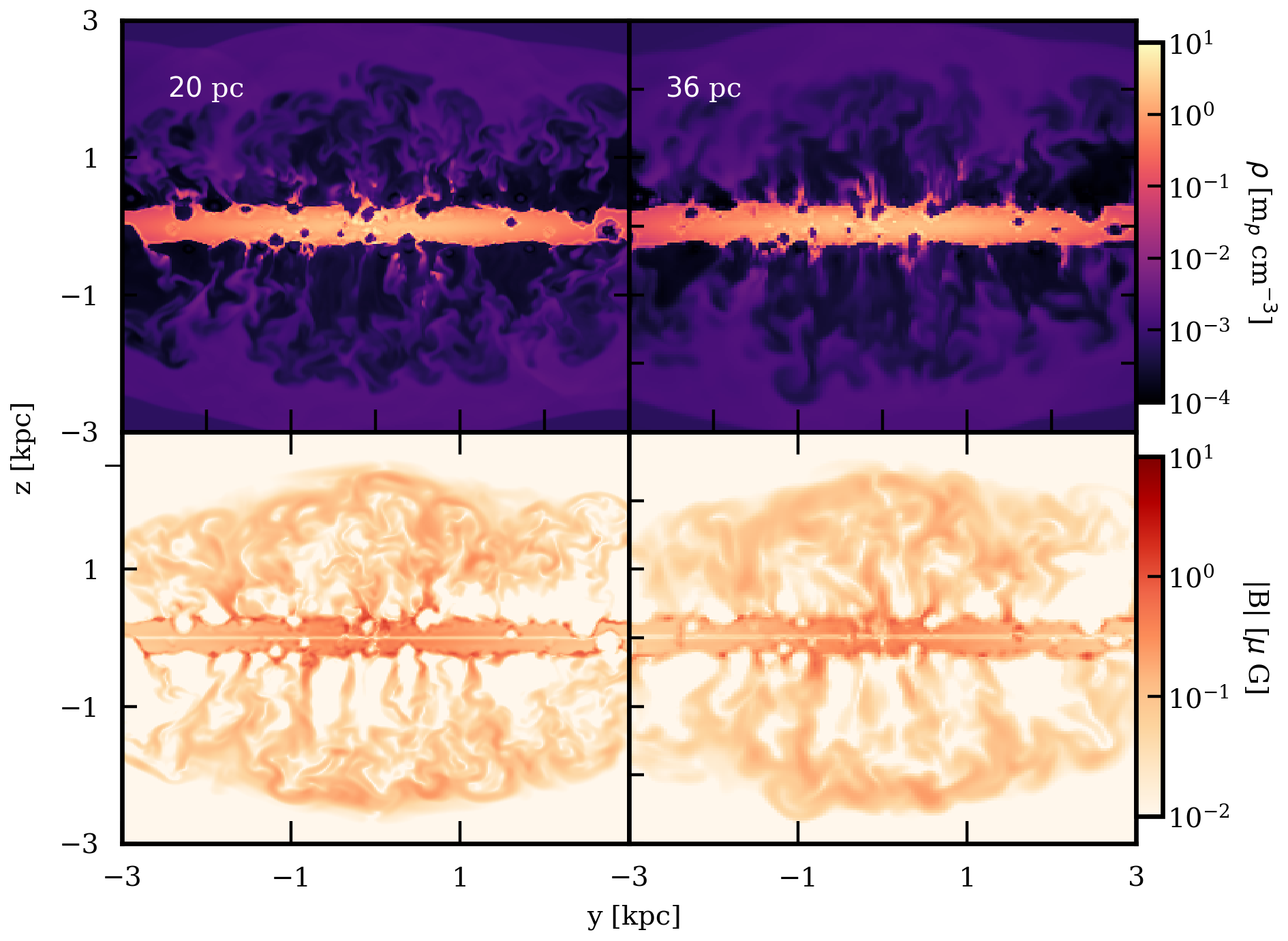}
	\caption{Slices in the $x=0$ plane showing density (top) and magnitude of magnetic field, $|B|=\sqrt{B_x^2 + B_y^2 + B_z^2}$, (bottom) at $5$ Myr. The right panel shows the results from simualations with resolution $20$ pc, while the left one is from $36$ pc, which corresponds to the fiducial run. Both the slices have similar morphology.}
	\label{fig:cong_slice}
\end{figure*}

\begin{figure}
    \centering
    \includegraphics[width=\columnwidth]{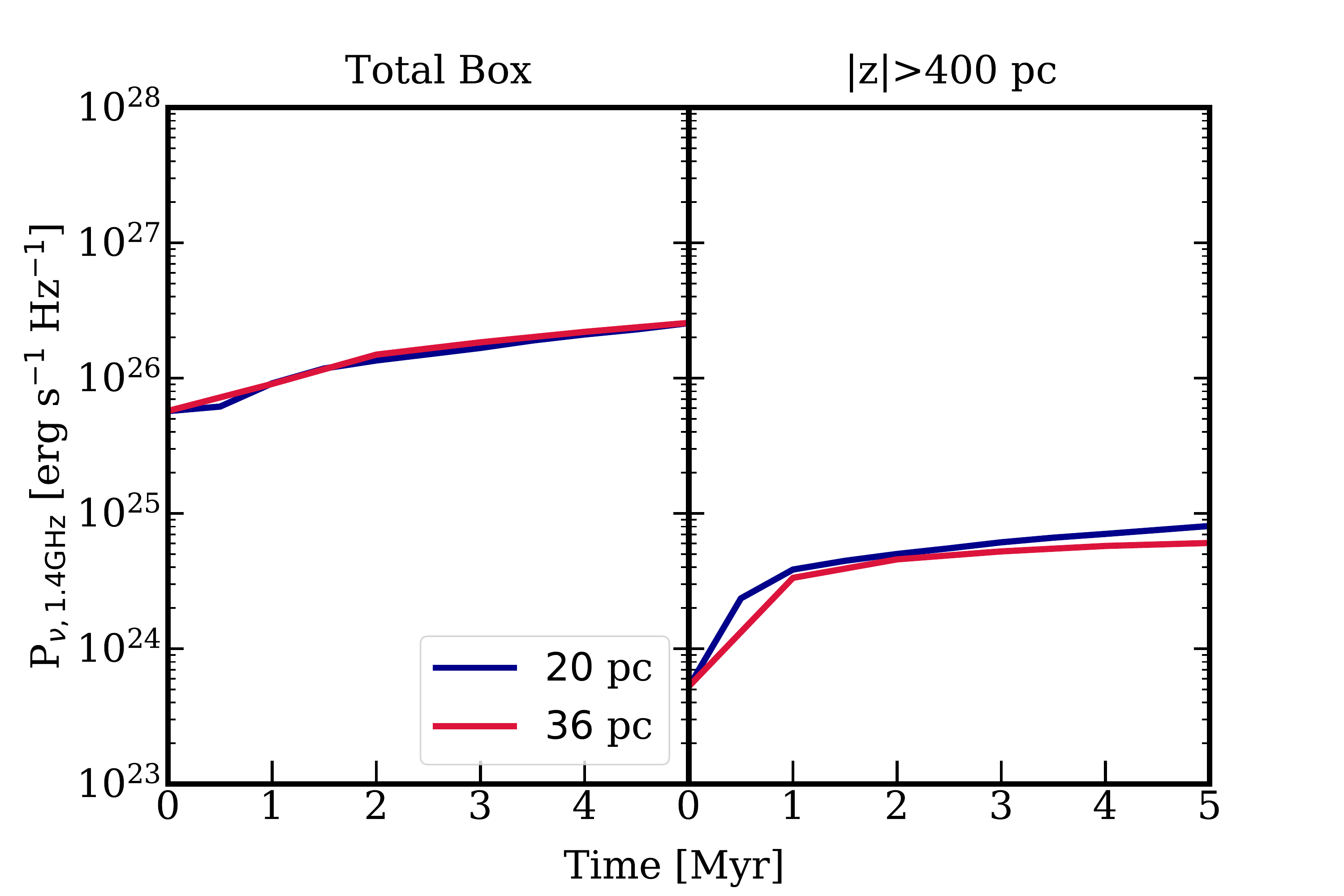}
	\caption{Evolution of total synchrotron power for SFR = $3$ \sfr at two different resolutions- $36$ pc (blue cuvre) and $50$ pc (red). The left panel shows the power emitted by the entire simulation domain. The right shows the power emitted by the extra-planar region.  }
	\label{fig:cong_totL}
\end{figure}

We conduct convergence tests for our simulations. We use resolutions of $20$ pc and compare it with our fiducial run at $36$ pc (SFR $= 3$ \sfr). We conduct the higher resolution with identical initial conditions, including the locations of OB associations. Since the high resolution run is computationally expensive, we limit the simulation domain of this run to $[-3,3]^3$ kpc$^3$. We compare the results from this run with a box of identical dimensions of the lower resolution run.

In Figure \ref{fig:cong_slice} we show density (top) and magnetic field magnitude (bottom) slices in $x=0$ plane for the two resolutions.  The general morphology in both the runs is similar for density as well magnetic field magnitude slices. From the density (top) panels, we note that the outer shock has reached a height of $z \sim 3$ kpc in $5$ Myr for both the cases. We note that a higher resolution run has sharper contact discontinuity at about $2$ kpc.  

In Figure \ref{fig:cong_totL} we compare the total synchrotron emitted as a function of time upto $5$ Myr, after which the outer shock moves out of the simulation domain for the higher resolution run. The left panel shows the total power emitted at $1.4$ GHz in the entire simulation domain. In the right panel we show the power for the entire simulation domain for $20$ pc (blue) run and $[-3,3]^3$ kpc$^3$ box for $36$ pc run (red). We see that power for the two resolutions matches closely in the left panel. In the right panel, we show total power from extra-planar gas. There is greater difference between power ($\sim 10 \%$). From Figure \ref{fig:zbrit}, we know that thermal energy density dominates in the extra-planar region over the kinetic energy. Because of higher resolution, in the extra-planar region the mixing between the cold gas and the hot gas is efficient, which raises the average temperature of the gas. 

From Figures \ref{fig:cong_slice} and \ref{fig:cong_totL}, we conclude that our simulations have converged.

\end{document}